\documentclass{PoS}
\usepackage{axodraw}

\def\lsim{\mathrel {\vcenter {\baselineskip 0pt \kern 0pt
    \hbox{$<$} \kern 0pt \hbox{$\sim$} }}}
\def\gsim{\mathrel {\vcenter {\baselineskip 0pt \kern 0pt
    \hbox{$>$} \kern 0pt \hbox{$\sim$} }}}
\newcommand{\la}[1]{\label{#1}}
\newcommand{\ba}{\begin{eqnarray}}
\newcommand{\ea}{\end{eqnarray}}
\newcommand{\rmi}[1]{{\mbox{\scriptsize #1}}}
\newcommand{\fig}{Fig.~}
\newcommand{\eq}{Eq.~}
\newcommand{\se}{Sec.~}
\newcommand{\eqs}{Eqs.~}
\newcommand{\nr}[1]{(\ref{#1})}
\newcommand{\tr}{{\mbox{Tr}\,}}
\newcommand{\nn}{\nonumber \\}
\renewcommand{\vec}[1]{{\bf #1}}
\newcommand{\lk}{\left[}
\newcommand{\rk}{\right]}
\newcommand{\Tc}{T_{\rm c}}
\newcommand{\Nf}{N_{\rm f}}
\newcommand{\Nc}{N_{\rm c}}
\newcommand{\bmu}{\bar\mu}
\newcommand{\rmii}[1]{{\mbox{\tiny\rm{#1}}}}

\newcommand{\mE}{m_\rmii{E}}
\newcommand{\gE}{g_\rmii{E}}
\newcommand{\lE}{\lambda_\rmii{E}}

\newcommand{\gM}{g_\rmii{M}}
\newcommand{\gammaE}{\gamma_\rmii{E}}
\newcommand{\rmO}{{\cal O}}
\newcommand{\Tint}[1]{{\hbox{$\sum$}\!\!\!\!\!\!\!\int\,}_{\!\!\!\!\raise-0.9ex\hbox{$\scriptstyle{#1}$}}}

\newcommand{\sy}[3]{{\scriptstyle #1\frac{#2}{#3}}}
\newcommand{\sm}[1]{\,{\scriptstyle #1}\,}

\def\Lwidth{1}

\def\Agl(#1,#2)(#3,#4,#5){\PhotonArc(#1,#2)(#3,#4,#5){\Lwidth}
{6.283 #3 mul 360 div #4 #5 sub #4 #5 sub mul sqrt mul Ldensity mul}}
\def\Lgl(#1,#2)(#3,#4){\Photon(#1,#2)(#3,#4){\Lwidth}
{#1 #3 sub #1 #3 sub mul #2 #4 sub #2 #4 sub mul add sqrt Ldensity mul}}
\def\Asc(#1,#2)(#3,#4,#5){\CArc(#1,#2)(#3,#4,#5)}
\def\Lsc(#1,#2)(#3,#4){\Line(#1,#2)(#3,#4)}
\def\Legl(#1,#2)(#3,#4){\Photon(#1,#2)(#3,#4){\Lwidth}
{#1 #3 sub #1 #3 sub mul #2 #4 sub #2 #4 sub mul add sqrt Ldensity mul}}

\def\scfc{0.7}  
\def\phgt{21}   
\def\pwcb{31.5} 
\newcommand{\PIC}[4]{\;\parbox[c]{#2 pt}{\begin{picture}(#2,#3)(0,0)
\SetWidth{1.0}\SetScale{#4} #1 \end{picture}}\;}

\newcommand{\picb}[1]{\PIC{#1}{\pwcb}{\phgt}{\scfc}}

\def\ToprSBB(#1,#2,#3,#4,#5){\picb{#1(0,15)(7.5,15)  #1(37.5,15)(45,15)%
 #2(22.5,15)(15,0,70) #2(22.5,15)(15,110,180) #3(22.5,15)(15,180,360)%
 #4(22.5,30)(5,-10,190) #5(22.5,30)(5,190,350)}}
\def\ToprSBT(#1,#2,#3,#4){\picb{#1(0,15)(7.5,15)  #1(37.5,15)(45,15)%
 #2(22.5,15)(15,0,90) #2(22.5,15)(15,90,180) #3(22.5,15)(15,180,360)%
 #4(22.5,35)(5,-90,270)}}
\def\ToprSTB(#1,#2,#3,#4){\picb{#1(0,0)(22.5,0) #1(22.5,0)(45,0)%
 #2(22.5,15)(15,-90,70) #2(22.5,15)(15,110,270)%
 #3(22.5,30)(5,-10,190) #4(22.5,30)(5,190,350)}}
\def\ToprSTT(#1,#2,#3){\picb{#1(0,0)(22.5,0) #1(22.5,0)(45,0)%
 #2(22.5,15)(15,-90,90) #2(22.5,15)(15,90,270)%
 #3(22.5,35)(5,-90,270)}}
\def\ToptSMx(#1,#2,#3,#4,#5,#6){\picb{#1(0,15)(7.5,15) #1(37.5,15)(45,15)%
 #2(22.5,15)(15,0,90) #3(22.5,15)(15,90,180) #4(22.5,15)(15,180,270)%
 #5(22.5,15)(15,270,360) #6(22.5,30)(22.5,0)%
 \GCirc(11.9,25.6){2.5}{0}}}
\def\ToptSMy(#1,#2,#3,#4,#5,#6){\picb{#1(0,15)(7.5,15) #1(37.5,15)(45,15)%
 #2(22.5,15)(15,0,90) #3(22.5,15)(15,90,180) #4(22.5,15)(15,180,270)%
 #5(22.5,15)(15,270,360) #6(22.5,30)(22.5,0)%
 \GCirc(22.5,15){2.5}{0}}}
\def\ToptSMz(#1,#2,#3,#4,#5,#6){\picb{#1(0,15)(7.5,15) #1(37.5,15)(45,15)%
 #2(22.5,15)(15,0,90) #3(22.5,15)(15,90,180) #4(22.5,15)(15,180,270)%
 #5(22.5,15)(15,270,360) #6(22.5,30)(22.5,0)%
 \GCirc(33.1,25.6){2.5}{0}}}
\def\ToptSMa(#1,#2,#3,#4,#5,#6){\picb{#1(0,15)(7.5,15) #1(37.5,15)(45,15)%
 #2(22.5,15)(15,0,90) #3(22.5,15)(15,90,180) #4(22.5,15)(15,180,270)%
 #5(22.5,15)(15,270,360) #6(22.5,30)(22.5,0)%
 \GCirc(7.5,15){2.5}{0}}}
\def\ToptSMb(#1,#2,#3,#4,#5,#6){\picb{#1(0,15)(7.5,15) #1(37.5,15)(45,15)%
 #2(22.5,15)(15,0,90) #3(22.5,15)(15,90,180) #4(22.5,15)(15,180,270)%
 #5(22.5,15)(15,270,360) #6(22.5,30)(22.5,0)%
 \GCirc(22.5,30){2.5}{0}}}
\def\ToprSBBx(#1,#2,#3,#4,#5){\picb{#1(0,15)(7.5,15)  #1(37.5,15)(45,15)%
 #2(22.5,15)(15,0,70) #2(22.5,15)(15,110,180) #3(22.5,15)(15,180,360)%
 #4(22.5,30)(5,-10,190) #5(22.5,30)(5,190,350)%
 \GCirc(11.9,25.6){2.5}{0}}}
\def\ToprSBBy(#1,#2,#3,#4,#5){\picb{#1(0,15)(7.5,15)  #1(37.5,15)(45,15)%
 #2(22.5,15)(15,0,70) #2(22.5,15)(15,110,180) #3(22.5,15)(15,180,360)%
 #4(22.5,30)(5,-10,190) #5(22.5,30)(5,190,350)%
 \GCirc(22.5,0){2.5}{0}}}
\def\ToprSBBz(#1,#2,#3,#4,#5){\picb{#1(0,15)(7.5,15)  #1(37.5,15)(45,15)%
 #2(22.5,15)(15,0,70) #2(22.5,15)(15,110,180) #3(22.5,15)(15,180,360)%
 #4(22.5,30)(5,-10,190) #5(22.5,30)(5,190,350)%
 \GCirc(22.5,25){2.5}{0}}}
\def\ToprSBBw(#1,#2,#3,#4,#5){\picb{#1(0,15)(7.5,15)  #1(37.5,15)(45,15)%
 #2(22.5,15)(15,0,70) #2(22.5,15)(15,110,180) #3(22.5,15)(15,180,360)%
 #4(22.5,30)(5,-10,190) #5(22.5,30)(5,190,350)%
 \GCirc(22.5,35){2.5}{0}}}
\def\ToprSBBa(#1,#2,#3,#4,#5){\picb{#1(0,15)(7.5,15)  #1(37.5,15)(45,15)%
 #2(22.5,15)(15,0,70) #2(22.5,15)(15,110,180) #3(22.5,15)(15,180,360)%
 #4(22.5,30)(5,-10,190) #5(22.5,30)(5,190,350)%
 \GCirc(7.5,15){2.5}{0}}}
\def\ToprSBBb(#1,#2,#3,#4,#5){\picb{#1(0,15)(7.5,15)  #1(37.5,15)(45,15)%
 #2(22.5,15)(15,0,70) #2(22.5,15)(15,110,180) #3(22.5,15)(15,180,360)%
 #4(22.5,30)(5,-10,190) #5(22.5,30)(5,190,350)%
 \GCirc(17.5,29){2.5}{0}}}
\def\ToptSAlx(#1,#2,#3,#4,#5){\picb{#1(0,15)(7.5,15) #1(37.5,15)(45,15)%
 #2(22.5,15)(15,0,90) #3(22.5,15)(15,90,180) #4(22.5,15)(15,180,360)%
 #5(7.5,30)(15,270,360)%
 \GCirc(11.9,25.6){2.5}{0}}}
\def\ToptSAlz(#1,#2,#3,#4,#5){\picb{#1(0,15)(7.5,15) #1(37.5,15)(45,15)%
 #2(22.5,15)(15,0,90) #3(22.5,15)(15,90,180) #4(22.5,15)(15,180,360)%
 #5(7.5,30)(15,270,360)%
 \GCirc(33.1,25.6){2.5}{0}}}
\def\ToptSAly(#1,#2,#3,#4,#5){\picb{#1(0,15)(7.5,15) #1(37.5,15)(45,15)%
 #2(22.5,15)(15,0,90) #3(22.5,15)(15,90,180) #4(22.5,15)(15,180,360)%
 #5(7.5,30)(15,270,360)%
 \GCirc(22.5,0){2.5}{0}}}
\def\ToptSAlw(#1,#2,#3,#4,#5){\picb{#1(0,15)(7.5,15) #1(37.5,15)(45,15)%
 #2(22.5,15)(15,0,90) #3(22.5,15)(15,90,180) #4(22.5,15)(15,180,360)%
 #5(7.5,30)(15,270,360)%
 \GCirc(17.2,20.3){2.5}{0}}}
\def\ToptSAla(#1,#2,#3,#4,#5){\picb{#1(0,15)(7.5,15) #1(37.5,15)(45,15)%
 #2(22.5,15)(15,0,90) #3(22.5,15)(15,90,180) #4(22.5,15)(15,180,360)%
 #5(7.5,30)(15,270,360)%
 \GCirc(22.5,30){2.5}{0}}}
\def\ToptSAlb(#1,#2,#3,#4,#5){\picb{#1(0,15)(7.5,15) #1(37.5,15)(45,15)%
 #2(22.5,15)(15,0,90) #3(22.5,15)(15,90,180) #4(22.5,15)(15,180,360)%
 #5(7.5,30)(15,270,360)%
 \GCirc(37.5,15){2.5}{0}}}
\def\ToptSAlc(#1,#2,#3,#4,#5){\picb{#1(0,15)(7.5,15) #1(37.5,15)(45,15)%
 #2(22.5,15)(15,0,90) #3(22.5,15)(15,90,180) #4(22.5,15)(15,180,360)%
 #5(7.5,30)(15,270,360)%
 \GCirc(7.5,15){2.5}{0}}}
\def\ToprSTBx(#1,#2,#3,#4){\picb{#1(0,0)(22.5,0) #1(22.5,0)(45,0)%
 #2(22.5,15)(15,-90,70) #2(22.5,15)(15,110,270)%
 #3(22.5,30)(5,-10,190) #4(22.5,30)(5,190,350)%
 \GCirc(7.5,15){2.5}{0}}}
\def\ToprSTBy(#1,#2,#3,#4){\picb{#1(0,0)(22.5,0) #1(22.5,0)(45,0)%
 #2(22.5,15)(15,-90,70) #2(22.5,15)(15,110,270)%
 #3(22.5,30)(5,-10,190) #4(22.5,30)(5,190,350)%
 \GCirc(22.5,25){2.5}{0}}}
\def\ToprSTBz(#1,#2,#3,#4){\picb{#1(0,0)(22.5,0) #1(22.5,0)(45,0)%
 #2(22.5,15)(15,-90,70) #2(22.5,15)(15,110,270)%
 #3(22.5,30)(5,-10,190) #4(22.5,30)(5,190,350)%
 \GCirc(22.5,35){2.5}{0}}}
\def\ToprSTBa(#1,#2,#3,#4){\picb{#1(0,0)(22.5,0) #1(22.5,0)(45,0)%
 #2(22.5,15)(15,-90,70) #2(22.5,15)(15,110,270)%
 #3(22.5,30)(5,-10,190) #4(22.5,30)(5,190,350)%
 \GCirc(17.5,29){2.5}{0}}}
\def\ToprSTBc(#1,#2,#3,#4){\picb{#1(0,0)(22.5,0) #1(22.5,0)(45,0)%
 #2(22.5,15)(15,-90,70) #2(22.5,15)(15,110,270)%
 #3(22.5,30)(5,-10,190) #4(22.5,30)(5,190,350)%
 \GCirc(22.5,0){2.5}{0}}}
\def\ToptSSx(#1,#2,#3,#4){\picb{#1(0,15)(7.5,15) #1(37.5,15)(45,15)%
 #4(7.5,15)(37.5,15) #2(22.5,15)(15,0,180) #3(22.5,15)(15,180,360)%
 \GCirc(22.5,15){2.5}{0}}}
\def\ToptSSy(#1,#2,#3,#4){\picb{#1(0,15)(7.5,15) #1(37.5,15)(45,15)%
 #4(7.5,15)(37.5,15) #2(22.5,15)(15,0,180) #3(22.5,15)(15,180,360)%
 \GCirc(22.5,30){2.5}{0}}}
\def\ToptSSc(#1,#2,#3,#4){\picb{#1(0,15)(7.5,15) #1(37.5,15)(45,15)%
 #4(7.5,15)(37.5,15) #2(22.5,15)(15,0,180) #3(22.5,15)(15,180,360)%
 \GCirc(7.5,15){2.5}{0}}}
\def\ToprSBTc(#1,#2,#3,#4){\picb{#1(0,15)(7.5,15)  #1(37.5,15)(45,15)%
 #2(22.5,15)(15,0,90) #2(22.5,15)(15,90,180) #3(22.5,15)(15,180,360)%
 #4(22.5,35)(5,-90,270)%
 \GCirc(22.5,30){2.5}{0}}}
\def\ToprSTTc(#1,#2,#3){\picb{#1(0,0)(22.5,0) #1(22.5,0)(45,0)%
 #2(22.5,15)(15,-90,90) #2(22.5,15)(15,90,270)%
 #3(22.5,35)(5,-90,270)%
 \GCirc(22.5,30){2.5}{0}}}
\def\ToptSEc(#1,#2,#3,#4,#5){\picb{#1(0,15)(7.5,15) #1(37.5,15)(45,15)%
 #3(15,15)(7.5,0,180) #4(15,15)(7.5,180,360)%
 #2(30,15)(7.5,0,180) #5(30,15)(7.5,180,360)%
 \GCirc(22.5,15){2.5}{0}}} 
\def\FiveA(#1,#2,#3,#4){\picb{#1(0,0)(22.5,0) #1(22.5,0)(45,0)%
 #2(22.5,15)(15,-90,90) #3(22.5,15)(15,90,270)#4(22.5,0)(22.5,30)%
 \GCirc(22.5,0){2.5}{0}}}
\def\FiveB(#1,#2,#3,#4){\picb{#1(0,15)(7.5,15) #1(37.5,15)(45,15)%
 #4(16,15)(8.5,0,360) #2(22.5,15)(15,0,180) #3(22.5,15)(15,180,360)%
 \GCirc(7.5,15){2.5}{0}}}
\def\SixA(#1,#2,#3){\picb{#1(0,15)(22.5,15) #1(22.5,15)(45,15)%
 #2(22.5,23.5)(8.5,0,360) #3(22.5,6.5)(8.5,0,360)%
 \GCirc(22.5,15){2.5}{0}}} 

\def\ToprS(#1){\SPropCircProp(#1,3)}
\def\ToprSMi(#1,#2,#3,#4,#5,#6,#7,#8){\picb{#1(0,15)(7.5,15) #1(37.5,15)(45,15)%
 #2(22.5,15)(15,0,90) #3(22.5,15)(15,90,180) #4(22.5,15)(15,180,270)%
 #5(22.5,15)(15,270,360) #6(22.5,30)(22.5,16) #7(22.5,16)(11.5,5) #8(22.5,16)(33.5,5)}}
\def\ToprSMiV(#1,#2,#3,#4,#5,#6,#7){\picb{#1(0,15)(7.5,15) #1(37.5,15)(45,15)%
 #2(22.5,15)(15,0,90) #3(22.5,15)(15,90,180) #4(22.5,15)(15,180,270)%
 #5(22.5,15)(15,270,360) #6(22.5,0)(11.5,25) #7(22.5,0)(33.5,25)}}
\def\ToprSMiNp(#1,#2,#3,#4,#5,#6,#7,#8){\picb{#1(0,15)(7.5,15) #1(37.5,15)(45,15)%
 #2(22.5,15)(15,0,90) #3(22.5,15)(15,90,180) #4(22.5,15)(15,180,270)%
 #5(22.5,15)(15,270,360) #6(33.5,5)(11.5,25) #7(26,16)(33.5,25) #8(11.5,5)(20,13)}}
\def\ToprSMiB(#1,#2,#3,#4,#5,#6){\picb{#1(0,15)(7.5,15) #1(37.5,15)(45,15)%
 #2(22.5,15)(15,0,90) #3(22.5,15)(15,90,180) #4(22.5,15)(15,180,360)%
 #5(7.5,30)(15,270,360) #6(37.5,0)(15,90,180)}}
\def\ToprSMiBT(#1,#2,#3,#4,#5,#6,#7){\picb{#1(0,15)(7.5,15) #1(37.5,15)(45,15)%
 #2(22.5,15)(15,0,90) #3(22.5,15)(15,90,180) #4(22.5,15)(15,180,360)%
 #5(22.5,30)(22.5,15) #6(22.5,15)(22.5,0) #7(22.5,15)(37,15)}}
\def\ToprSMiBB(#1,#2,#3,#4,#5,#6,#7){\picb{#1(0,15)(7.5,15) #1(37.5,15)(45,15)%
 #2(22.5,15)(15,0,90) #3(22.5,15)(15,90,180) #4(22.5,15)(15,180,360)%
 #5(22.5,15)(5,0,360) #6(22.5,30)(22.5,20) #7(22.5,10)(22.5,0)}}

\newcommand{\picA}{\begin{picture}(50,40)%
\Oval(25,20)(20,25)(0)\Line(25,0)(10,36)\Line(25,0)(40,36)%
\put(50,20){\circle*{4}}\end{picture}}
\newcommand{\picB}{\begin{picture}(50,40)%
\Oval(25,20)(20,25)(0)\Line(25,0)(10,36)\Line(25,0)(40,36)%
\put(25,40){\circle*{4}}\end{picture}}
\newcommand{\picC}{\begin{picture}(50,40)%
\Oval(25,20)(20,25)(0)\Line(25,0)(10,36)\Line(25,0)(40,36)%
\put(25,40){\circle*{4}}\put(50,20){\circle*{4}}%
\Line(-2,18)(2,22)\Line(-2,22)(2,18)\end{picture}}
\newcommand{\picD}{\begin{picture}(50,40)%
\Oval(25,20)(20,25)(0)\Line(25,0)(10,36)\Line(25,0)(40,36)%
\put(20,39.5){\circle*{4}}\put(30,39.5){\circle*{4}}%
\Line(48,18)(52,22)\Line(48,22)(52,18)\end{picture}}
\newcommand{\picE}{\begin{picture}(50,40)%
\Oval(25,20)(20,25)(0)\Line(25,0)(10,36)\Line(25,0)(40,36)%
\put(25,40){\circle*{4}}\put(17,39){\circle*{4}}\put(33,39){\circle*{4}}%
\Line(-2,18)(2,22)\Line(-2,22)(2,18)\Line(48,18)(52,22)\Line(48,22)(52,18)%
\end{picture}}
\newcommand{\picF}{\begin{picture}(50,40)%
\Oval(25,20)(20,25)(0)\Oval(25,20)(20,10)(0)%
\put(2,28){\circle*{4}}\put(2,12){\circle*{4}}\end{picture}}

\newcommand{\picu}[1]{\;\parbox[c]{40pt}{\begin{picture}(50,30)(0,0)
\SetWidth{1.0}\SetScale{0.5} #1 \end{picture}}\; }
\def\DiagA{\picu{%
 \SetWidth{2.0} 
 \PhotonArc(30,30)(15,0,360){1.5}{15}%
 \Photon(45,30)(60,30){1.5}{3}%
 \Photon(34.64,44.27)(39.27,58.53){1.5}{3}%
 \Photon(17.86,38.82)(5.73,47.63){1.5}{3}%
 \Photon(17.86,21.18)(5.73,12.37){1.5}{3}%
 \Photon(34.64,15.73)(39.27,1.47){1.5}{3}%
}}
\def\DiagB{\picu{%
 \SetWidth{2.0} 
 \DashArrowArc(30,30)(15,0,360){2}%
 \Photon(45,30)(60,30){1.5}{3}%
 \Photon(34.64,44.27)(39.27,58.53){1.5}{3}%
 \Photon(17.86,38.82)(5.73,47.63){1.5}{3}%
 \Photon(17.86,21.18)(5.73,12.37){1.5}{3}%
 \Photon(34.64,15.73)(39.27,1.47){1.5}{3}%
}}
\def\DiagC{\picu{%
 \SetWidth{2.0} 
 \PhotonArc(30,30)(15,0,360){1.5}{15}%
 \Photon(45,30)(57.12,38.82){1.5}{3}%
 \Photon(45,30)(57.12,21.18){-1.5}{3}%
 \Photon(30,45)(30,60){1.5}{3}%
 \Photon(14,30)(-1,30){1.5}{3}%
 \Photon(30,15)(30,0){1.5}{3}%
}}
\def\DiagD{\picu{%
 \SetWidth{2.0} 
 \DashArrowArc(30,30)(15,-45,315){2}%
 \Photon(45,30)(57.12,38.82){1.5}{3}%
 \Photon(45,30)(57.12,21.18){-1.5}{3}%
 \Photon(30,45)(30,60){1.5}{3}%
 \Photon(15,30)(0,30){1.5}{3}%
 \Photon(30,15)(30,0){1.5}{3}%
}}
\def\DiagE{\picu{%
 \SetWidth{2.0} 
 \PhotonArc(30,30)(15,0,360){1.5}{15}%
 \Photon(45,30)(60,30){1.5}{3}%
 \Photon(22.5,42.99)(26.38,57.48){1.5}{3}%
 \Photon(22.5,42.99)(8.01,46.87){-1.5}{3}%
 \Photon(22.5,17.01)(26.38,2.52){-1.5}{3}%
 \Photon(22.5,17.01)(8.01,13.13){1.5}{3}%
}}
\def\DiagF{\picu{%
 \SetWidth{2.0} 
 \DashArrowArc(30,30)(15,0,360){2}%
 \Photon(45,30)(60,30){1.5}{3}%
 \Photon(22.5,42.99)(26.38,57.48){1.5}{3}%
 \Photon(22.5,42.99)(8.01,46.87){-1.5}{3}%
 \Photon(22.5,17.01)(26.38,2.52){-1.5}{3}%
 \Photon(22.5,17.01)(8.01,13.13){1.5}{3}%
}}

\title{3-loop gauge coupling for hot gauge theories}

\ShortTitle{Hot gauge coupling}

\author{\speaker{York Schr\"oder}
\\Grupo de Cosmolog\'ia y Part\'iculas Elementales, Universidad del B\'io-B\'io, Chill\'an, Chile
\\E-mail: \email{yschroder@ubiobio.cl}}

\abstract{This talk offers a brief review of the determination of coupling constants in the framework of dimensionally reduced effective field theories for thermal QCD, specializing on its gluonic sector. Interestingly, higher-order operators that go beyond the usual super-renormalizable truncation of the effective theory need to be considered when matching parameters at three loops.}

\FullConference{Loops and Legs in Quantum Field Theory (LL2018)\\29 April 2018 - 04 May 2018\\St. Goar, Germany}

\begin{document}


\section{Introduction}

When studying the physics of a hot system of strongly interacting matter (where we have mainly quarks and gluons in mind), dimensionally reduced thermal effective theories prove most valuable for a systematic understanding of effects originating from separate energy scales \cite{dr1,dr2,generic}. 
In this brief review, we will focus on equilibrium thermodynamics of quantum chromodynamics (QCD), where interesting questions such as for example the study of phase transitions connected to confinement and chiral symmetry breaking, can be addressed. 
Answers to these questions are relevant for a wide spectrum of phenomenological applications, such as in the fields of compact star astrophysics or early-universe cosmology, or in heavy-ion collision experiments that probe a quark-gluon plasma (QGP).

At very large temperatures, asymptotic freedom asserts the existence of a theoretically tractable limit of QCD, where the gauge coupling becomes small and weak-coupling methods are applicable. This opens the prospect of first-principles studies of this QCD regime, with analytic methods and possibility of systematic improvements, without the need to resort to models. Besides the temperature $T$, other dimensionful parameters of the system can be quark chemical potentials $\mu_q$ and quark-masses $m_q$, as well as dimensionless characteristics of the fermion and gauge representations ($\Nf$ and $\Nc$). We will focus here on a pure gauge $SU(\Nc)$ theory only, and hence be concerned with dependence on only $T$ and $\Nc$. 

Close to a phase transition (or crossover), such as the deconfinement-confinement transition at the critical temperature $\Tc$ of the order of 175 MeV, the QGP is strongly coupled and has to be treated by non-perturbative methods, such as e.g.\ lattice Monte Carlo simulations. However, as already mentioned above, at $T\gg\Tc$ a weak-coupling approach can be used. One caveat is that, as has been pointed out long ago \cite{linde}, a strict loop expansion is not well-defined, due to infrared (IR) divergences at higher loop orders. While we will see below why and how this matters, and how IR effects can be systematically accounted for, let us remark here that in general, one uses a mix of discrete (lattice) and continuum (perturbative) methods, each one where it works best, in order to make predictions over a sizable energy interval. Again, the focus here will be on the continuum side.

At high $T$ (and/or $\mu_q$), interactions make QCD a multi-scale system. Indeed, the expansion parameter is not simply the strong coupling constant $\alpha_s\sim g^2$, where $g$ denotes the gauge coupling parameter, but it gets multiplied by a (bosonic) distribution function that accounts for the multiple interactions (with typical momentum $k$, say) in the gluon-plasma: 
$ g^2\,n_b(|k|) = \frac{g^2}{e^{|k|/T}-1}$. At asymptotically high $T$ we have $g\ll1$, such that three momentum scales can be cleanly separated. Parametrically, these are of the order $|k|\sim \{T, g T, g^2 T\}$; they correspond to the typical momentum scale of particles in a heatbath of temperature $T$, and to dynamically generated mass-scales for the two gluon polarizations; they are conventionally called `hard', `soft' and `ultrasoft' scales; and, as can be seen by expanding the Bose function at small $|k|/T$ as $n_b(|k|)\approx T/|k|$, induce expansion parameters that are of order $g^2$ ($g$) for hard (soft) modes, but of order unity for the ultrasoft ones, rendering the latter non-perturbative even at high temperatures. Due to confinement-like behavior in the ultrasoft sector, there are no smaller momentum scales. Hence, this multi-scale system with three well separated (at high $T\gg\Tc$ or equivalently $g\ll1$) scales allows for a most transparent treatment in terms of effective field theory (EFT), as will be made precise below.


\section{Effective theory setup}
\la{se:eft}

At high temperatures, the dynamics of QCD is contained in a 3-dimensional (3d) effective theory \cite{gpy,nadkarni}, which is conventionally called `electrostatic QCD' (EQCD) \cite{bn}. One essentially `integrates out' the effects of the hard momentum scales, $|\vec{k}|\gsim\pi T$, keeping only field components with smaller (soft and ultrasoft) momenta dynamical. In this manner, one integrates out the quark fields completely (since they do not possess a Matsubara zero mode and are hence `heavy' spectators), and one integrates out the non-zero Matsubara modes of the gauge fields, ending up with the purely bosonic EQCD with Lagrangian
\ba\la{eq:eqcd}
 {\cal L}_\rmi{EQCD} &=& -\frac1{2\gE^2} \tr [D_i,D_j]^2 + \tr [D_k,A_0]^2 + 
 \mE^2\,\tr A_0^2 +\lE^{(1)} (\tr A_0^2)^2
 +\lE^{(2)}\, \tr A_0^4 + \dots \;.
\ea
Here, $A_0$ is a scalar (the notation specifying it as a remnant of the 4-dimensional (4d) gauge field $A_\mu$), and the 3d gauge field sits in the covariant derivative $D_k=\partial_k-i\gE A_k$. All fields $A_0,\,A_k$ are in the fundamental representation of $SU(\Nc)$. In \eq\nr{eq:eqcd}, we have not shown the gauge fixing term and have omitted higher-order operators; for the latter, see \se\ref{se:dim6} below. Note that in 3d, the gauge coupling acquires a mass dimension, see e.g.\ \eq\nr{eq:gEpar}.

The parameters are of course not arbitrary, but fixed in terms of the parameters of the parent theory, 4d QCD. 
They can be determined systematically by requiring weak-coupling expansions of a set of $n$\/-point functions to coincide for scales where both descriptions (4d QCD and 3d EQCD) hold, resulting in perturbative expressions such as
\ba
\la{eq:gEpar}
\gE^2 &=& T\lk g^2+n_0\,g^4+n_1\,g^6+n_2\,g^8+\dots\rk \;,\\
\mE^2 &=& T^2\lk n_3\,g^2+n_4\,g^4+n_5\,g^6+\dots\rk \;,\la{eq:m}\\
\lE^{(1),(2)} &=& T\lk n_6\,g^4+n_7\,g^6+\dots\rk \;.
\ea
The determination of the coefficients $n_i$ has a long history; 2- and 3-loop results have been presented in \cite{gE2,ig_mE,ig,ig_gE}.
All $n_i$ shown above are known analytically, except for $n_2$ on which we will report in \se\ref{se:3} below.

Having performed the QCD $\rightarrow$ EQCD reduction, one realizes immediately that ${\cal L}_\rmi{EQCD}$ contains a mass term for the scalar $A_0$, which calls for another reduction step. Indeed, integrating out effects of the soft momentum scales, $|\vec{k}|\gsim gT$, allows to systematically eliminate $A_0$ (whose mass according to \eq\nr{eq:m} is $\mE\sim gT$), ending up with a 3d pure gauge theory, which is conventionally called `magnetostatic QCD' (MQCD). It is defined by
\ba
\la{eq:mqcd}
{\cal L}_\rmi{MQCD} &=& -\frac1{2\gM^2}\tr[D_i,D_j]^2+\dots \;,
\ea
where we have again not shown higher-order operators beyond the superrenormalizable ones (they will be briefly discussed in \se\ref{se:us} below), and where the covariant derivative now contains the gauge coupling $\gM$,  which can be determined in terms of the parameters of the parent theory, 3d EQCD, as \cite{pg2,gE2}
\ba
\gM^2 &=& \gE^2 \lk1+n_8\,\frac{\gE^2}{\mE}+n_9\,\frac{\gE^4}{\mE^2}+\dots\rk\;.
\ea
According to the effective theory setup QCD$\rightarrow$EQCD$\rightarrow$MQCD as sketched above, IR effects are now captured by 3d MQCD.


\section{Determination of matching coefficients: 3-loop gauge coupling}
\la{se:3}

Let us now discuss how the parameters of the effective theory, or matching coefficients, are determined in practice, with precision, and in a systematically improvable manner. For $\mE$ and the $\lE$ we refer to the literature \cite{ig_mE,gE2}, but discuss the effective gauge coupling $\gE$ in more detail here. 


\subsection{Hard contributions}
\la{se:hard}

When reducing QCD to EQCD, an efficient way to determine matching coefficients is to shift the gauge field $A\rightarrow A+B$ and evaluate the effective action for a background field $B$ \cite{abbott}.
For example, the effective gauge coupling parameter $\gE^2 = g^2/[{\cal Z}_B+\delta {\cal Z}_B]$ can then be read from the quadratic part of the background-field effective action 
$\Gamma_\rmi{EQCD}^{(2)}[B]=\frac12 B^a_i(p)B^a_j(q)\delta(p+q)(q^2\delta_{ij}-q_i q_j)[{\cal Z}_B+\delta{\cal Z}_B]$.
The main ingredient can hence be seen to be the transverse part of the two-point function, or background-field self-energy; in particular, we need the second term of its expansion around small external momenta, $\Pi_T'(0)$ \cite{gE2}. 
 
\begin{figure}[t]

\vspace*{-2mm}

\begin{eqnarray*}
\sm{+1} \ToprSMi(\Legl,\Agl,\Agl,\Agl,\Agl,\Lgl,\Lgl,\Lgl)
\sm{+1} \ToprSMiV(\Legl,\Agl,\Agl,\Agl,\Agl,\Lgl,\Lgl)
\sy+14 \ToprSMiNp(\Legl,\Agl,\Agl,\Agl,\Agl,\Lgl,\Lgl,\Lgl)
\sy+14 \ToprSMiB(\Legl,\Agl,\Agl,\Agl,\Agl,\Agl)
\sy+14 \ToprSMiBB(\Legl,\Agl,\Agl,\Agl,\Agl,\Lgl,\Lgl)
\sy+12 \ToprSMiBT(\Legl,\Agl,\Agl,\Agl,\Lgl,\Lgl,\Lgl)
\sm{+\;\;441\;\mbox{diags}}
\end{eqnarray*}

\vspace*{-2mm}

\caption[a]{\small 
Typical 3-loop contributions to the 2-point function in the background field gauge.
The diagrams have been drawn with the help of Axodraw~\cite{axodraw}.}
\la{fig:2pt}
\end{figure}

An evaluation of the required Feynman diagrams up to three loops starts with a chain of standard computer algebra tools and algorithms, adapted to 4d thermal field theories. 
In a first step, diagrams are generated with {\tt QGRAF} \cite{qgraf}, resulting in $\sim 450$ two-point diagrams at 3 loops, some representatives of which are shown in \fig\ref{fig:2pt}. 
Secondly, symbolic manipulation in {\tt FORM} \cite{form} projects the calculation onto $\sim 10^7$ vacuum sum-integrals.
Third, systematic use of linear integration-by-parts (IBP) relations \cite{lapo} applied to the 3d piece of the sum-integrals achieves a reduction to $\sim 10^2$ so-called `master' sum-integrals, of which $\sim 10^1$ are bosonic \cite{jm}. 
Using the IBP tables, a basis transformation of the bosonic masters can be performed in order to render the actual polynomial pre-factors of non-trivial masters finite as $d\rightarrow4$, such that it suffices to evaluate them up to their constant parts. The structure of the resulting set of six non-trivial bosonic 3-loop master sum-integrals is depicted in 
\fig\ref{fig:mas}.

Turning to the evaluation of the master sum-integrals, we remind the reader that at finite temperatures we have a compact (imaginary) time interval that leads to (Matsubara) sums in momentum space, whence the integral measure is
\ba
\Tint{P}=T\!\!\sum_{n=-\infty}^{\infty}
\int\frac{d^{3-2\epsilon}p}{(2\pi)^{3-2\epsilon}} \;.
\ea
Massless propagators are then $\frac1{P^2}=\frac1{P_0^2+\vec p^2}$ with (in the bosonic case that we are discussing here) $P_0=2\pi n T$.
While 1-loop massless vacuum sum-integrals are trivial (they evaluate to Zeta values) and at two loops factorize into products of 1-loop cases, they start to be nasty objects starting at 3 loops.
The evaluation of masters such as those shown in \fig\ref{fig:mas} is a highly non-trivial task, as no standard algorithmic methods are known for higher-loop sum-integrals. 
As a consequence, most available 3-loop results are very specific cases that utilize the specific (spectacle- and basketball-type) structures of the cases at hand, exploiting their 1-loop sub-structure. Pioneering work has been done in the nineties \cite{az}, based on which many other beautiful methods have been developed, such as e.g.\ lifting Tarasov's T-operators \cite{Top} to finite temperature \cite{tensor} in order to trade tensors for dimension shifts. 
The general strategy is a careful dissection of the sum-integral into divergent and finite pieces, with divergences evaluated analytically and finite parts containing numerical results, the transcendental number content of which is still an interesting open question. 
To appreciate the structure of such results for 3-loop sum-integrals, we refer the reader to consult some of our earlier contributions to this conference series, where the first and last of the six masters of \fig\ref{fig:mas} have been presented \cite{ll10,ll12}.

\begin{figure}[t]
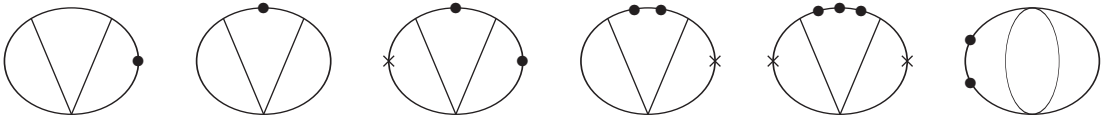


\vspace*{-2mm}

\begin{eqnarray*}
\picA\qquad
\picB\qquad
\picC\qquad
\picD\qquad
\picE\qquad
\picF
\end{eqnarray*}

\vspace*{-3mm}

\caption[a]{\small 
The six non-trivial bosonic 3-loop master sum-integrals that we need for the hard contribution to $\gE^2$. A single line corresponds to a massless propagator $1/P^2$; a line with a cross carries an extra numerator factor $P_0^2$; and a line decorated with one (two, three) dots stands for the 2nd (3rd, 4th) power of a massless propagator, respectively.}
\la{fig:mas}
\end{figure}

Finally, after accounting for gauge coupling and wave function renormalization \cite{pert,contlatt}, one arrives at the renormalized NNLO result for effective gauge coupling \cite{ig,ig_gE} (from above, recall the representation $\gE^2 \;=\; g^2/[{\cal Z}_B+\delta {\cal Z}_B+\rmO(g^8)]$)
\ba
{\cal Z}_B &=&  
1 \!-\!G \bigg[ \frac{22}{3} L \!+\! \frac{1}{3} \bigg]
\!-\!G^2 \bigg[ \frac{68}{3} L \!+\! \frac{341}{18} \!-\! \frac{10}{9} \zeta_3 \bigg]
\!-\! G^3 \bigg[ \frac{748}{9} L^2
\!+\!\bigg(  \frac{6608}{27} \!-\! \frac{10982}{135} \zeta_3 \!\bigg) L
\!+\! \mbox{(finite)} \bigg] \;,\nn
\la{eq:div}
\delta {\cal Z}_B &=& G^3 \bigg[ \frac{61\zeta_3}{5\epsilon} \bigg]
\;,\qquad G\;=\;\frac{g^2 \Nc}{16\pi^2}
\;,\qquad L\;=\;\ln \biggl( \frac{\bmu e^{\gammaE}}{4\pi T} \biggr) \;.
\ea
Interestingly, the result contains a remaining $1/\epsilon$ divergence, which we have separated in $\delta {\cal Z}_\rmi{B}$.
The origin (and cure) of this remaining logarithmic divergence will be investigated in what follows.


\subsection{Dimension-six operators in EQCD}
\la{se:dim6}

To shine some light on the fate of the divergence in \eq\nr{eq:div}, let us now examine higher-order operators in the effective theory that have been omitted from the EQCD Lagrangian. The dimension-six operators that can be added to \eq\nr{eq:eqcd} have been classified in \cite{chapman}, albeit in strictly 4d, which (in view of the fact that we have divergences and work in dimensional regularization) we have to generalize to general $d$. Their structure is
\ba\la{eq:Deqcd}
 \delta {\cal L}_\rmi{EQDC} &=& 
 \bigg(\Tint{P}' \frac{2 \gE^2 }{P^6}\bigg) \; \mbox{tr}
 \Bigl\{ 
  c^{ }_1 \, (D^{ }_\mu F^{ }_{\mu\nu})^2 + 
  c^{ }_2 \, (D^{ }_\mu F^{ }_{\mu 0})^2 
 \nn 
 && +i\gE 
 \bigl[
  c^{ }_3\, F^{ }_{\mu\nu} F^{ }_{\nu\rho} F^{ }_{\rho\mu } + 
  c^{ }_4\, F^{ }_{0 \mu} F^{ }_{\mu\nu} F^{ }_{\nu 0  } + 
  c^{ }_5\, A^{ }_0 (D^{ }_{\mu} F^{ }_{\mu\nu}) F^{ }_{ 0 \nu}  
 \bigr]
 \\
 && 
 +\gE^2 
 \bigl[ 
  c^{ }_6\, A_0^2 F_{\mu\nu}^2 + 
  c^{ }_7 \, A^{ }_0 F^{ }_{\mu\nu} A^{ }_0 F^{ }_{\mu\nu} + 
  c^{ }_8 \, A_0^2 F_{0\mu}^2 +
  c^{ }_9 \, A^{ }_0 F^{ }_{0 \mu} A^{ }_0 F^{ }_{0\mu} 
 \bigr] 
 + \gE^4 
 \bigl[
  c^{ }_{10} A_0^6 \,
 \bigr]
 \Bigr\} \;, \nonumber
\ea
where, to facilitate comparison with \cite{chapman}, we have now written the color traces in the adjoint representation, $\mbox{tr}(AB)=A_{ab}B_{ba}$ with e.g.\ $(A_0)_{ab}=-if^{abc}A_0^c$ etc. The operator basis is actually non-minimal, as there is one linear relation between $c_4,\dots,c_7$; we keep this redundancy for crosschecks. 
The 1-loop sum-integral in the first line of \eq\nr{eq:Deqcd}, where the prime on the sum excludes the Matsubara zero mode $n=0$, evaluates to Gamma and Zeta functions and is finite,
\ba
\Tint{P}' \frac{T^2}{P^6} &=& \frac{2\zeta_3}{(4\pi)^4} \bigg[1+{\cal O}(\epsilon)\bigg] \:.
\ea

\begin{figure}[t]
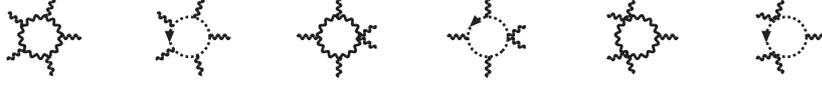


\vspace*{-2mm}

\hspace*{0.2cm}%
\begin{minipage}[c]{14cm}
\begin{eqnarray*}
 && 
 \DiagA \quad 
 \DiagB \quad
 \DiagC \quad
 \DiagD \quad
 \DiagE \quad
 \DiagF 
\end{eqnarray*}
\end{minipage}

\caption[a]{\small 
Contributions to the 5-point function in the background field gauge. Here, wiggly and dotted lines stand for gluons and ghosts, respectively.}
\la{fig:5pt}
\end{figure}

For perturbative expansions involving the operators of \eq\nr{eq:Deqcd}, we clearly need their coefficients $c_i$ in $d$ dimensions. This has been done in \cite{soft}; one can evaluate e.g.\ 1-loop contributions to the 5-point function, see \fig\ref{fig:5pt}, which contains 20 independent Lorentz structures that allow to fix the LO results for the $c_i$ as (restricting to Feynman gauge here)
\ba
\la{eq:ci}
&&c^{ }_1 =  \frac{41-d}{120}
\;,\;\; c^{ }_2 =  \frac{(d-1)(d-5)}{120}
\;,\;\; c^{ }_3 =  \frac{1-d}{180}
\;,\;\; c^{ }_4 - 2 c^{ }_7 = \frac{(41-d)(5-d)}{60}\;,
\\&&
c^{ }_5 - 2 c^{ }_7 = \frac{(21\!-\!d)(5\!-\!d)}{30}
\;,\;\; c^{ }_6 + c^{ }_7 = \frac{(d\!-\!25)(5\!-\!d)}{24}
\;,\;\; c^{ }_8 = \frac{(5\!-\!d)(3\!-\!d)(d\!-\!1)}{20}\;,
\\&&
c^{ }_9 = \frac{(5-d)(3-d)(d-1)}{30}
\;,\;\; c^{ }_{10} = \frac{(5-d)(3-d)(d-1)^2}{180}\;.
\ea
To check these expressions, the 2-, 3- and 6-point functions have been evaluated as well in \cite{soft}, finding full agreement.
Recalling that EQCD is defined in $d=3-2\epsilon$ dimensions, we see that $c_8$, $c_9$ and $c_{10}$ couple to `evanescent' operators (and were therefore not accounted for in \cite{chapman}). 


\subsection{Soft contributions}

What is now the effect of considering the dimension-six operators of $\delta {\cal L}_\rmi{EQDC}$?
When integrating out the soft scales, i.e.\ reducing EQCD $\rightarrow$ MQCD, one needs to determine the MQCD gauge coupling $\gM^2$. In full analogy to the above, it is convenient to determine $\gM^2=\gE^2/[Z_B+\delta Z_B]$ from 2-point functions in background-field gauge.
Note that, since we are dealing with a 3d computation here, we need at least two loops to see a logarithmic divergence. The corresponding diagrams containing some of the new vertices arising from $\delta {\cal L}_\rmi{EQDC}$ are shown in \fig\ref{fig:dim6b}, where graphs with closed loops of massless lines have already been omitted.

\begin{figure}[t]
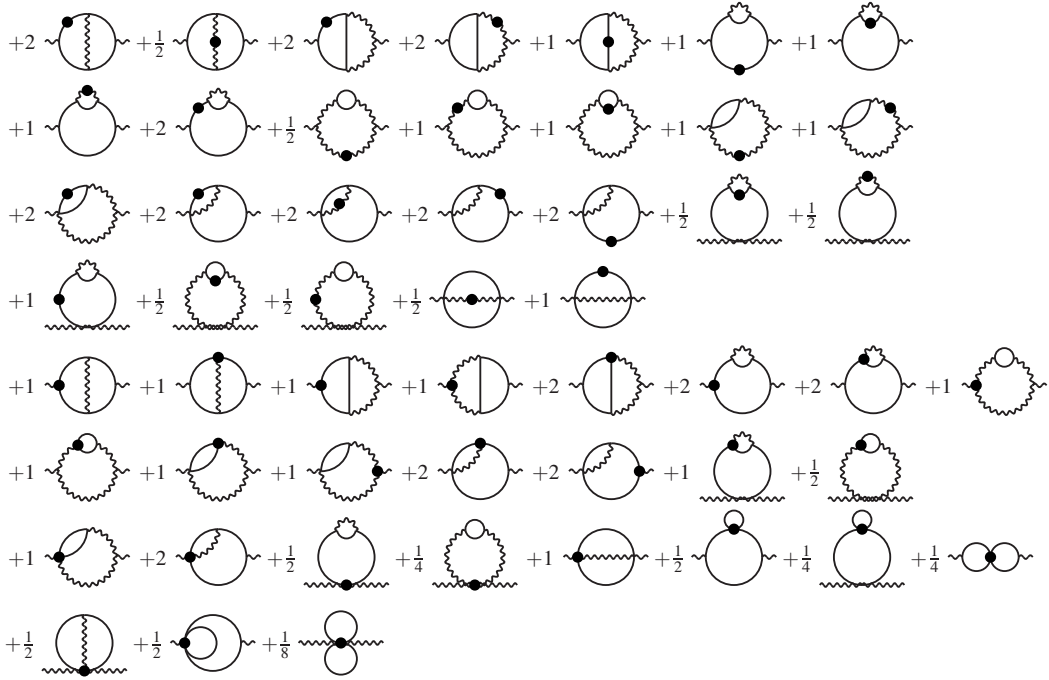


\vspace*{-2mm}

\hspace*{0.1cm}%
\begin{minipage}[c]{15.0cm}
\begin{eqnarray*}
&&{} \hspace*{-0.3cm}
\sm{+2} \ToptSMx(\Legl,\Asc,\Asc,\Asc,\Asc,\Lgl) 
\sy+12  \ToptSMy(\Legl,\Asc,\Asc,\Asc,\Asc,\Lgl) 
\sm{+2} \ToptSMx(\Legl,\Agl,\Asc,\Asc,\Agl,\Lsc)
\sm{+2} \ToptSMz(\Legl,\Agl,\Asc,\Asc,\Agl,\Lsc)
\sm{+1} \ToptSMy(\Legl,\Agl,\Asc,\Asc,\Agl,\Lsc)
\sm{+1} \ToprSBBy(\Legl,\Asc,\Asc,\Agl,\Asc)
\sm{+1} \ToprSBBz(\Legl,\Asc,\Asc,\Agl,\Asc)
 \nn[1.5ex]&&{} \hspace*{-0.3cm}
\sm{+1} \ToprSBBw(\Legl,\Asc,\Asc,\Agl,\Asc)
\sm{+2} \ToprSBBx(\Legl,\Asc,\Asc,\Agl,\Asc)
\sy+12  \ToprSBBy(\Legl,\Agl,\Agl,\Asc,\Asc)
\sm{+1} \ToprSBBx(\Legl,\Agl,\Agl,\Asc,\Asc)
\sm{+1} \ToprSBBz(\Legl,\Agl,\Agl,\Asc,\Asc)
\sm{+1} \ToptSAly(\Legl,\Agl,\Asc,\Agl,\Asc)
\sm{+1} \ToptSAlz(\Legl,\Agl,\Asc,\Agl,\Asc)
 \nn[1.5ex]&&{} \hspace*{-0.3cm}
\sm{+2} \ToptSAlx(\Legl,\Agl,\Asc,\Agl,\Asc)
\sm{+2} \ToptSAlx(\Legl,\Asc,\Asc,\Asc,\Agl)
\sm{+2} \ToptSAlw(\Legl,\Asc,\Asc,\Asc,\Agl)
\sm{+2} \ToptSAlz(\Legl,\Asc,\Asc,\Asc,\Agl)
\sm{+2} \ToptSAly(\Legl,\Asc,\Asc,\Asc,\Agl)
\sy+12  \ToprSTBy(\Legl,\Asc,\Agl,\Asc)
\sy+12  \ToprSTBz(\Legl,\Asc,\Agl,\Asc)
 \nn[1.5ex]&&{} \hspace*{-0.3cm}
\sm{+1} \ToprSTBx(\Legl,\Asc,\Agl,\Asc)
\sy+12  \ToprSTBy(\Legl,\Agl,\Asc,\Asc) 
\sy+12  \ToprSTBx(\Legl,\Agl,\Asc,\Asc) 
\sy+12  \ToptSSx(\Lgl,\Asc,\Asc,\Lgl) 
\sm{+1} \ToptSSy(\Lgl,\Asc,\Asc,\Lgl) 
 \nn[1.5ex]&&{} \hspace*{-0.3cm}
\sm{+1} \ToptSMa(\Legl,\Asc,\Asc,\Asc,\Asc,\Lgl) 
\sm{+1} \ToptSMb(\Legl,\Asc,\Asc,\Asc,\Asc,\Lgl) 
\sm{+1} \ToptSMa(\Legl,\Agl,\Asc,\Asc,\Agl,\Lsc)
\sm{+1} \ToptSMa(\Legl,\Asc,\Agl,\Agl,\Asc,\Lsc)
\sm{+2} \ToptSMb(\Legl,\Agl,\Asc,\Asc,\Agl,\Lsc)
\sm{+2} \ToprSBBa(\Legl,\Asc,\Asc,\Agl,\Asc)
\sm{+2} \ToprSBBb(\Legl,\Asc,\Asc,\Agl,\Asc)
\sm{+1} \ToprSBBa(\Legl,\Agl,\Agl,\Asc,\Asc)
 \nn[1.5ex]&&{} \hspace*{-0.3cm}
\sm{+1} \ToprSBBb(\Legl,\Agl,\Agl,\Asc,\Asc)
\sm{+1} \ToptSAla(\Legl,\Agl,\Asc,\Agl,\Asc)
\sm{+1} \ToptSAlb(\Legl,\Agl,\Asc,\Agl,\Asc)
\sm{+2} \ToptSAla(\Legl,\Asc,\Asc,\Asc,\Agl)
\sm{+2} \ToptSAlb(\Legl,\Asc,\Asc,\Asc,\Agl)
\sm{+1} \ToprSTBa(\Legl,\Asc,\Agl,\Asc)
\sy+12  \ToprSTBa(\Legl,\Agl,\Asc,\Asc) 
 \nn[1.5ex]&&{} \hspace*{-0.3cm}
\sm{+1} \ToptSAlc(\Legl,\Agl,\Asc,\Agl,\Asc)
\sm{+2} \ToptSAlc(\Legl,\Asc,\Asc,\Asc,\Agl)
\sy+12  \ToprSTBc(\Legl,\Asc,\Agl,\Asc)
\sy+14  \ToprSTBc(\Legl,\Agl,\Asc,\Asc) 
\sm{+1} \ToptSSc(\Lgl,\Asc,\Asc,\Lgl) 
\sy+12 \ToprSBTc(\Legl,\Asc,\Asc,\Asc)
\sy+14 \ToprSTTc(\Legl,\Asc,\Asc) 
\sy+14 \ToptSEc(\Lgl,\Asc,\Asc,\Asc,\Asc)
 \nn[1.5ex]&&{} \hspace*{-0.3cm}
\sy+12 \FiveA(\Legl,\Asc,\Asc,\Lgl) 
\sy+12 \FiveB(\Legl,\Asc,\Asc,\Asc) 
\sy+18 \SixA(\Legl,\Asc,\Asc) 
\end{eqnarray*}
\end{minipage}

\vspace*{4mm}

\caption[a]{\small 
Contributions to the 2-loop 2-point function involving the new 2-, 3-, 4-, 5- and 6-point vertices arising from \eq\nr{eq:Deqcd}, which are denoted by a blob. Solid lines represent the massive adjoint scalar $A_0$.}
\la{fig:dim6b}
\end{figure}

After accounting for coupling and mass-renormalization, the next-to-leading order result for the 2-point correlator reads \cite{soft}
\ba
 Z_B & = & 
 1 \,+\, 
 G_\rmii{E}^2
 \frac{ \mE^{ }}{2\pi T}
 \biggl( \frac{875 \zeta^{ }_3 }{72} \biggr)
 \la{finalZ2}
 -G_\rmii{E}^3
 \biggl( \frac{1097 \zeta^{ }_3 }{549} \biggr)
 \frac{61}{5}
 \biggl\{ 
     L
 + 2 \ln\biggl( \frac{\bmu}{2 \mE^{ }} \biggr)
 + \frac{\zeta'_3}{\zeta^{ }_3}
 - \gammaE 
 + \frac{103771}{52656}   
 \biggr\}\;,\nn
\la{eq:div2}
 \delta Z_B & = & 
 G_\rmii{E}^3
 \bigg( \!\!-\frac{1097}{1098} \bigg)
 \bigg[\frac{61\zeta_3}{5\epsilon}\bigg] 
\;,\qquad G_\rmii{E}\;=\;\frac{\gE^2 \Nc}{16\pi^2 T}
\;,\qquad L\;=\;\ln \biggl( \frac{\bmu e^{\gammaE}}{4\pi T} \biggr) \;.
\ea
Taking into account the matching of the gauge couplings from \eq\nr{eq:gEpar} and 
comparing with \eq\nr{eq:div}, we note that this cancels $\frac{1097}{1098}$ of the IR divergence from the hard scales.


\subsection{Ultrasoft contributions}
\la{se:us}

So far, we have integrated out hard ($\sim T$) and soft ($\sim gT\sim\mE$) scales, and managed to understand a large fraction of the puzzling leftover divergence of \eq\nr{eq:div}. To proceed in the full spirit of the effective theory setup outlined in \se\ref{se:eft}, it clearly remains to check potential contributions from ultrasoft ($\sim g^2T$) scales.

The story of the preceding two sections repeats itself: classifying dimension-six operators of MQCD that had been omitted from \eq\nr{eq:mqcd}, one possible representation is \cite{soft}
\ba
 \delta {\cal L}_\rmi{MQCD} = 
 \bigg(\Tint{P}' \frac{2 \gM^2 }{P^6}\bigg) \; \mbox{tr}
 \Big\{ 
  c_1 \, (D_i F_{ij})^2
  + i\gM c_3\, F_{ij} F_{jk} F_{ki}  \Big\} \;,
\ea
where to leading order, the $c_i$ are proportional to those given in \eq\nr{eq:ci} \cite{cpa,soft}.
To extract UV divergences, it is sufficient here to screen the IR by a common (unphysical) mass, which results in 
\ba
\la{eq:div3}
\Tint{P}' \frac{\gM^6 \Nc^3}{P^6}\;\frac{T^2\,c_3}{32\pi^2\,\epsilon} \Big[1+{\cal O}(\epsilon)\Big] 
&\;\ni\;& \delta Z_{B} = G^3\,
 \bigg( -\frac{1}{1098} \bigg)
 \bigg[\frac{61\zeta_3}{5\epsilon}\bigg] \;,
\ea
where we have shown only the divergence (coming from 2-loop diagrams, recalling that MQCD is defined in 3d).

Finally, adding up \eqs\nr{eq:div}, \nr{eq:div2} and \nr{eq:div3}, the remaining logarithmic divergence cancels perfectly! 


\section{Conclusions}

Summarizing, for the 3-loop computation of the effective gauge coupling $\gE$, we indeed needed to consider dimension-six operators in both effective theories, EQCD and MQCD, in order to not miss any contributions from soft and ultrasoft momentum scales. In retrospect, this had been signaled by IR divergences of the hard sector, which duly cancel only after the full tower of effective theories has been considered.

A curious observation is that, although the soft scale $\mE\sim gT$ is formally larger than the ultrasoft scale $\sim g^2T$, it apparently plays an essential role in IR dynamics. In fact, comparing their respective impact (in terms of the IR divergence in the 3-loop gauge coupling; see \eqs\nr{eq:div2} and \nr{eq:div3}), the contribution of the ultrasoft scale is numerically dwarfed by that of the soft scale.

For future work, once all finite contributions to the 3-loop EQCD gauge coupling are available, one can envision an update of the comparison between 3d EFT- and 4d lattice-evaluations of the spatial string tension, as had been done at the 2-loop level in \cite{gE2,lat05}. This would serve as an important verification of the validity of the effective field theory framework, and provide motivation to generalize the Yang-Mills results discussed in \se\ref{se:hard} to full QCD.


\section*{Acknowledgments}

It is a pleasure to acknowledge the fruitful collaboration with M.~Laine and P.~Schicho on topics presented here, as well as their comments on the manuscript. 
I would also like to thank J.~M\"oller and I.~Ghisoiu for important contributions in early stages of the project.
This work was partly supported by FONDECYT project 1151281 and UBB project GI-172309/C.
I am grateful for support from the visitor program of the Albert Einstein Center for Fundamental Physics during a short-term stay at the University of Bern, where part of this work was done. 



\end{document}